\documentclass[prb,aps,twocolumn,showpacs]{revtex4}
\usepackage{graphicx}
\usepackage{wasysym}
\usepackage{bm}
\usepackage{pstricks}
\usepackage{subfigure}

\newcommand{\be}{\begin{equation}}
\newcommand{\ee}{\end{equation}}
\newcommand{\ba}{\begin{eqnarray}}
\newcommand{\ea}{\end{eqnarray}}
\newcommand{\n}{\nonumber \\}

\def\suhe#1{{\noindent\bf#1:}}

\begin{document}

\title{Classical Antiferromagnetism on Torquato-Stillinger Packings}

\author{F. J. Burnell$^1$ and S. L. Sondhi$^{1,2}$}

\affiliation{$^1$Department of Physics, Princeton University,
Princeton, NJ 08544, USA}
\affiliation{$^2$Princeton Center for Theoretical Physics, Princeton University,
Princeton, NJ 08544, USA}

\begin{abstract}
Torquato and Stillinger have constructed a new family of
frustrated lattices by unusually high dilution of close packed
structures while preserving structural stability. We show that an
infinite subclass of these structures has an underlying topology
that greatly simplifies determination of their magnetic phase
structure for nearest neighbor antiferromagnetism interactions and
$O(N)$ spins.

\end{abstract}
\maketitle

\section{Introduction} \label{intro}

The face centered cubic lattice and its close packed relatives are interesting
in two distinct contexts. The first is their structural stability as close
packed structures \cite{AM}. The second is their giving rise to geometrical frustration
when they host magnetic degrees of freedom \cite{fcc-mag, fcc-mag2}.

In a recent development, Torquato and Stillinger \cite{Torquato}
asked an intriguing question about these structures from the
stability viewpoint: how many sites can you dilute from them
without rendering them structurally unstable to shear forces?
Their answer is a family of packings, which we shall term
Torquato-Stillinger (TS) packings, with a local coordination
number of 7 and a spherical packing density of $\frac{\sqrt{2}
\pi}{9}$; for details we refer to Ref.~\onlinecite{Torquato}.

In this paper we examine the TS packings from the viewpoint of
geometrical frustration which survives the dilutions they
envisage. Specifically we determine the ground states, low
temperature ordering and the nature of the phase transition to the
paramagnetic state in a large subclass of the TS packings for
nearest neighbor antiferromagnetic interactions for classical
Ising, XY, and Heisenberg spins and as well as for $O(N)$ spins
with $N \ge 4$. In this task we will be greatly aided by the
simple observation that this subclass of TS packings is
topologically equivalent to a set of stacked triangular lattices
with half of the stacking bonds removed. As stacked triangular
lattices have been studied intensively (see
Refs.~\onlinecite{collinspetrenko}, \onlinecite{Loison} and
references therein), we will be able to carry over various results
from that work.

In the following we first review the TS construction of their packings and
single out a dilution of the FCC packing (lattice) as exemplifying the stacked triangular
structure that we will focus on. Next we present results for antiferromagnetism
on the TS-FCC packing and its relatives. We end with brief
comments on the cases not studied in this paper.

\section{Lattices}  \label{lattices}

General TS packings are constructed by removing one third of the
sites from a close packing of spheres.  To describe them more
precisely recall first that any three dimensional close packing of
spheres can be obtained by stacking two dimensionally close packed
triangular lattices of spheres according to a prescribed stacking
pattern. In a given triangular plane, the interior of triangular
plaquettes host depressions into which further spheres may be
placed. Of these we may select either all upward pointing
triangles or all downward pointing triangles in which to place the
spheres of the next triangular layer. The standard description
labels the original sites as, say, C whereupon the two
inequivalent depressions that host the second layer are labeled A
and B. All layers consist of spheres occupying one of these three
sets of sites with the rule that there is no repetition between
adjacent layers---this gives rise to the $2^N$ Barlow packings for
$N$ layers of spheres. As is well known, of these the repeated
sequences ABC yield the FCC lattice and AB or AC yield the hcp
structure \cite{AM}.

\begin{figure}[ht]
\begin{center}
\includegraphics[totalheight=5cm]{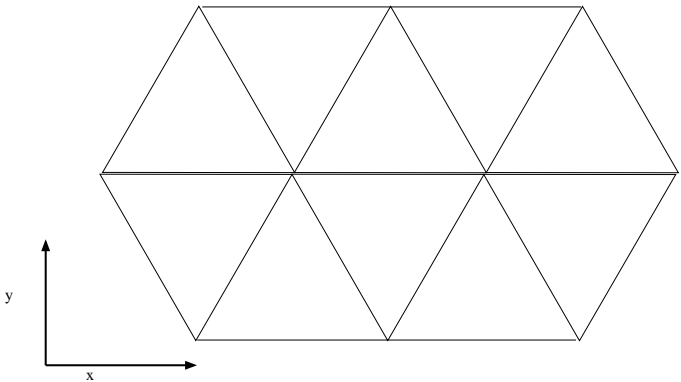}
\end{center}
\caption{\label{Triangles} The triangular lattice}
\end{figure}

An equivalent description can be given in terms of two {\it stacking vectors} which
allow us to translate one triangular layer into a neighboring one. As they can
be chosen independently at each step, we recover the previous counting. For
concreteness, let us orient one of the triangular layers as shown in Fig.
\ref{Triangles}. Now
the stacking vectors are readily seen to be
\ba
{\bf V}_\alpha &=& -\frac{a}{2}\hat{\bf x} +\frac{\sqrt{3}a}{6}\hat{\bf y} +
                   a\sqrt{\frac{2}{3}}\hat{\bf z} \n
{\bf V}_\beta &=& \frac{\sqrt{3}a}{3} \hat{\bf y} + a\sqrt{\frac{2}{3}} \hat{\bf z}
\ea
where $a$ is the diameter of the spheres. Now, for example, the allowed configurations
of three planes can be written as CAC (or ${\bf V}_\alpha$, ${\bf V}_\beta$),
CBC (${\bf V}_\beta$, ${\bf V}_\alpha$),
CAB (${\bf V}_\alpha$, ${\bf V}_\alpha$),  CBA (${\bf V}_\beta$, ${\bf V}_\beta$).

To dilute a sphere packing into a TS packing, one vertex of each
triangle in the triangular layers is removed, leaving stacked
honeycomb layers. Now at each step there are 6 choices---a choice
between two of the A, B, or C sites followed by a choice of which
of the three equivalent sublattices of the triangular layer to
dilute. All choices lead to stable structures \cite{Torquato}.

\begin{figure}[ht!]
\begin{center}
\includegraphics[totalheight=7cm]{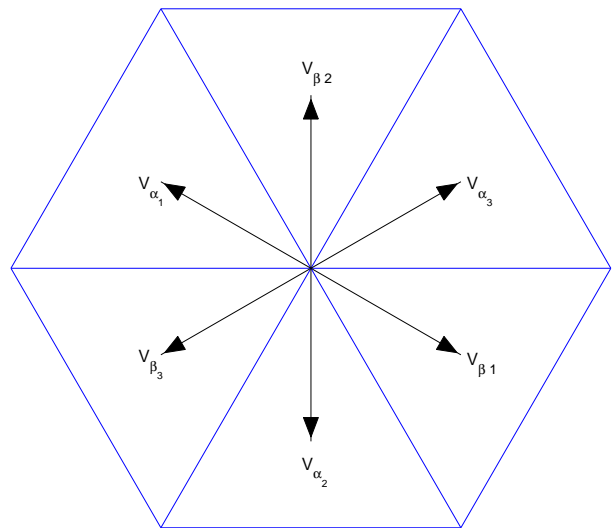}
\end{center}
\caption{\label{Stackings} Projections of the 6 stacking vectors in the honeycomb
  planes.  Note that all $V_{\alpha i}$ result in a CA stacking, and all $V_{\beta i}$ in a CB stacking. }
\end{figure}

Equivalently, we may begin with one honeycomb layer and construct
the rest of the structure by displacing it successively by
stacking vectors drawn now from a set of six vectors. With our
choice of orientation these are \ba {\bf V}_{\beta 1}&=&
\frac{a}{2} \hat{\bf x} -\frac{\sqrt{3}a}{6}\hat{\bf y} +
a\sqrt{\frac{2}{3}} \hat{\bf z} \n {\bf V}_{\beta 2} &=&
\frac{\sqrt{3}a}{3} \hat{\bf y} + a\sqrt{\frac{2}{3}} \hat{\bf z}
\n {\bf V}_{\beta 3} &=&  -\frac{a}{2}\hat{\bf x}
-\frac{\sqrt{3}a}{6}\hat{\bf y} + a\sqrt{\frac{2}{3}}\hat{\bf z}
\n {\bf V}_{\alpha 1}& =&   -\frac{a}{2}\hat{\bf x}
+\frac{\sqrt{3}a}{6}\hat{\bf y} + a\sqrt{\frac{2}{3}} \n {\bf
V}_{\alpha 2} &=& - \frac{\sqrt{3}a}{3} \hat{\bf y} +
a\sqrt{\frac{2}{3}} \hat{\bf z} \n {\bf V}_{\alpha 3} &=&
\frac{a}{2}\hat{\bf x}  +\frac{\sqrt{3}a}{6}\hat{\bf y} +
a\sqrt{\frac{2}{3}}\hat{\bf z} \n \ea Now, starting as above with
a C plane, ${\bf V}_{\alpha 1}$ through ${\bf V}_{\alpha 3}$
generate A planes, while ${\bf V}_{\beta 1}$ through ${\bf
V}_{\beta 3}$ yield B planes. The projections of these vectors in
the honeycomb planes are shown in Fig~\ref{Stackings}. Observe
that the projections of the ${\bf V}_{\beta i}$ are inverses of
the projections of the ${\bf V}_{\alpha i}$.

Three comments are in order. First, the TS packings contain tunnels through the parent
Barlow packings. The stacking vectors can also be visualized as giving the direction
of the tunnels (in other words, the offset between the centers of the
missing spheres in adjacent planes).  Second,
in the close packed case, irrespective of the stacking
pattern, rotations by $2 \pi /3$ radians about either the vertex or the
center of a triangle are symmetries of the structure. After removing
the centers of each hexagon, however, such rotations map some occupied
sites to unoccupied sites and vice versa, breaking the symmetry. Third, all
TS packings have a local coordination number of 7---3 nearest neighbors in a
single honeycomb layer and 2 each in the layer above and below.

Clearly, there are $6^N$ TS packings for $N$ honeycomb layers. In this article we
focus on a subset of them which is $2^N$ in number. We begin with one member of
this subset which is defined by the single stacking vector ${\bf V}_{\beta 1}$.  This
particular choice known as the tunneled FCC lattice, is discussed
extensively in Ref.~\onlinecite{Torquato}; we will review its structure briefly here.

\begin{figure}[htp]
\begin{center}
\subfigure[\label{TorLat-a}]{
\includegraphics[totalheight=5cm]{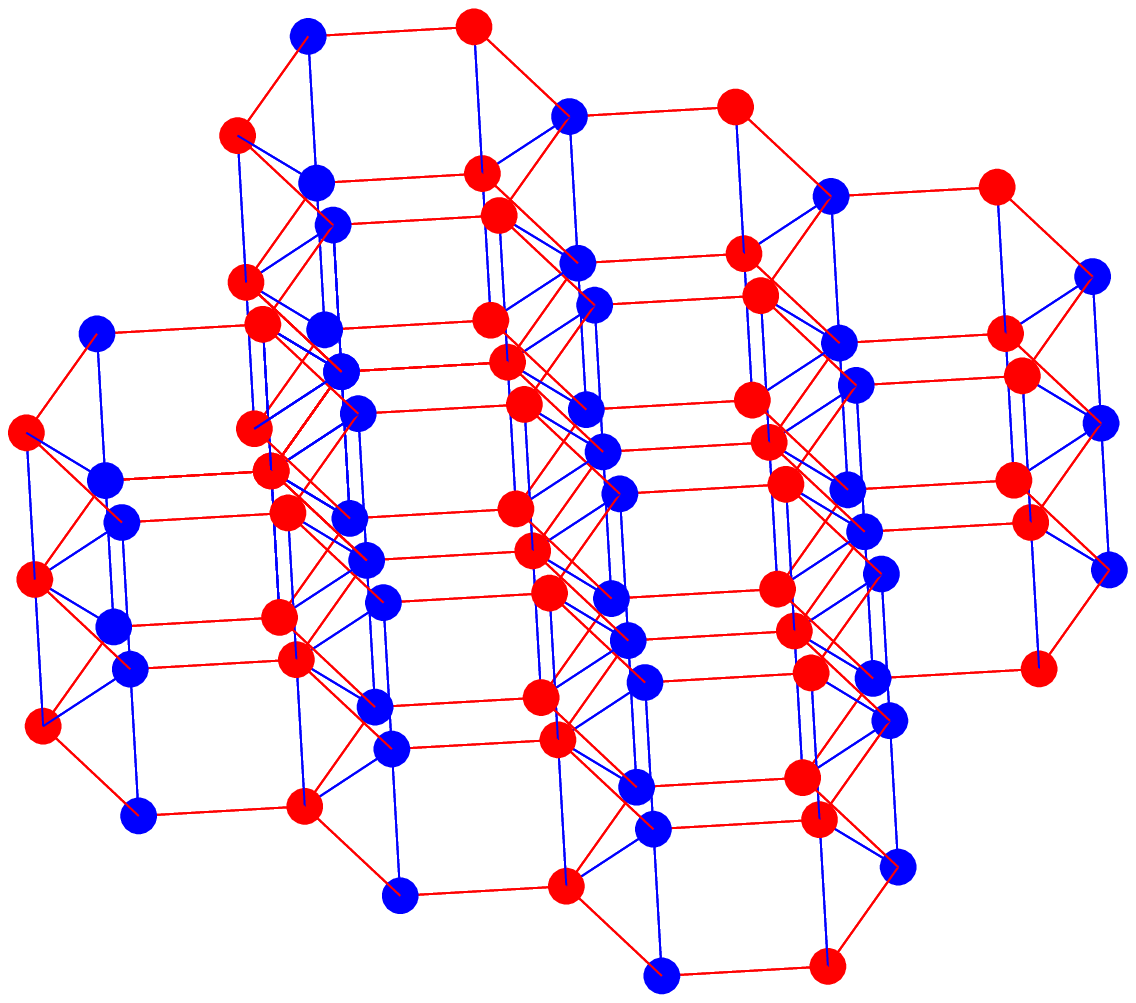}}
\subfigure[\label{TorLat-b}]{
\includegraphics[totalheight=5cm]{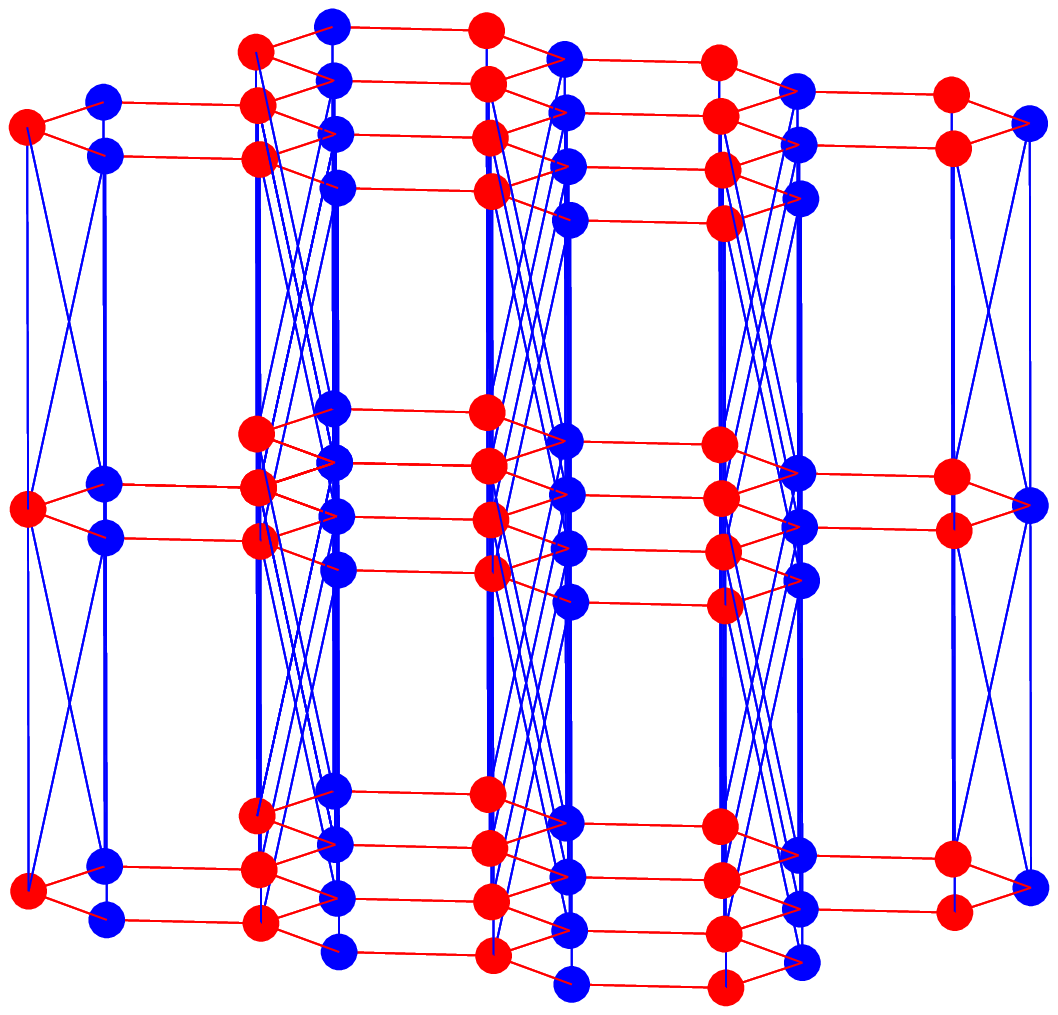}}
\subfigure[\label{TorLat-c}]{
\includegraphics[totalheight=5cm]{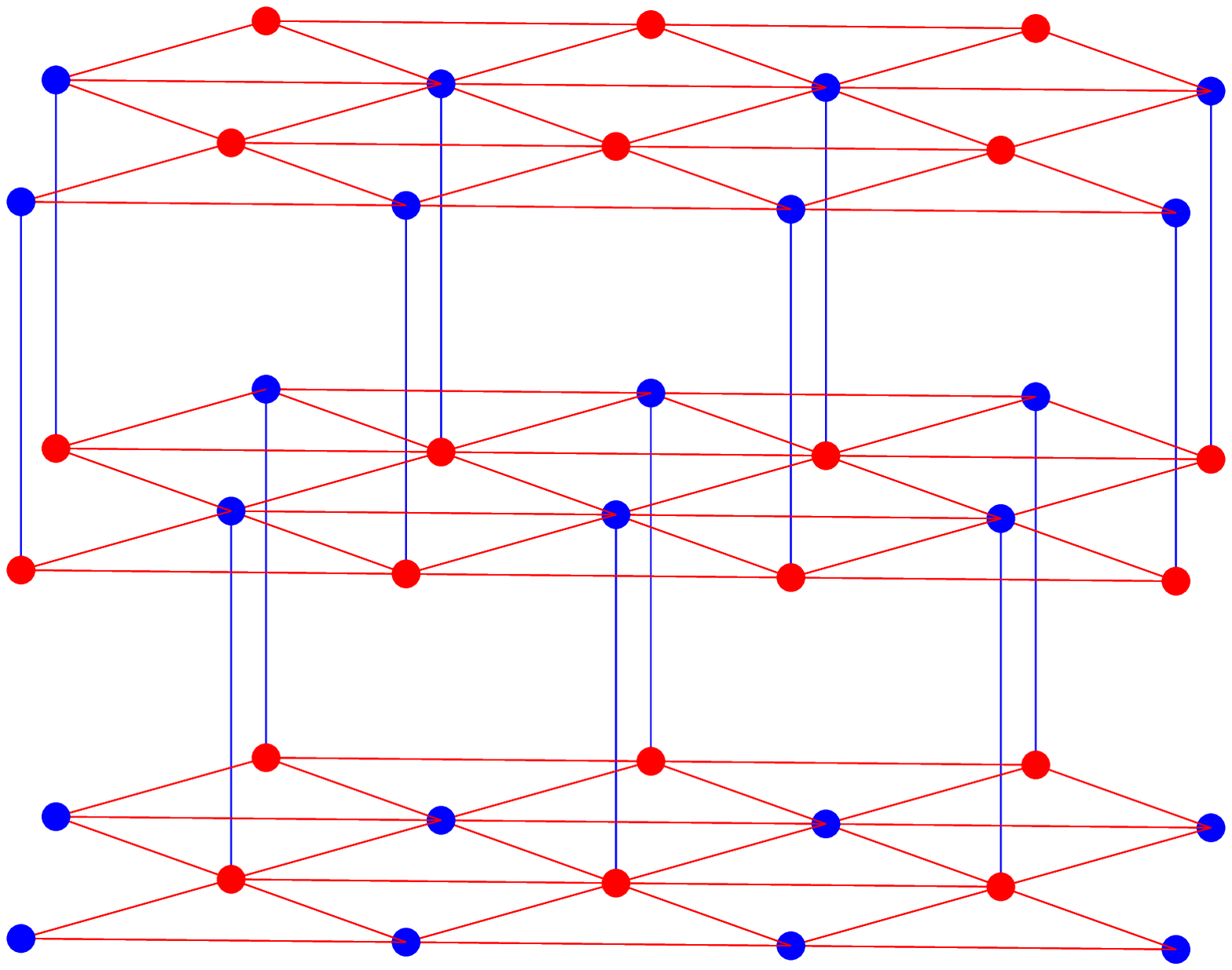}}
\end{center}
\caption{\label{TorLat} (a) The TS-FCC lattice as a set of stacked
honeycomb lattices.  Bonds in the honeycomb lattice ($xy$ plane)
are colored red (dotted lines); bonds joining different honeycomb
layers are blue (solid lines). The two colorings of the sites
differentiate the 2 sublattices. (b) A rotated view that exhibits
the alternate decomposition as a set of semi-stacked folded
triangular planes. The planes are seen almost edge on and consist
of sites from both sublattices. The lighter red (darker blue)
sites are connected to dark blue (light red) sites in the folded
triangular plane to the left (right). (c) The topologically
equivalent stacked triangular lattice, with unfolded triangular
planes now redrawn in the $xy$ plane.}
\end{figure}

Written conventionally, this packing is a triclinic
lattice with a two site unit cell.
The primitive lattice vectors are
\ba \label{primlatvecs}
{\bf a_1} &=& a (0, \sqrt{3}, 0) \n
{\bf a_2} &=& a ({3 \over 2}, -\frac{\sqrt{3}}{ 2},  0) \n
{\bf a_3} &=& a ({1 \over 2},- \frac{\sqrt{3}}{6},  \sqrt{2 \over 3} )
\ea
The two atoms of the unit cell are at positions
\ba \label{basisvecs}
{\bf x}_1 &=&  a (0,0,0) \n
{\bf x}_2 &=& a (1,0,0 )
\ea
The resulting lattice, shown in Fig.~\ref{TorLat-a}, consists of the
honeycomb lattice in the $xy$ plane, with nearest neighbours separated by
a distance $a$.  The honeycomb layers are stacked in the $z$ direction
according to the FCC pattern, with the same stacking vector ${\bf V}_{\beta 1}$
between every honeycomb plane.

We are primarily interested in nearest neighbor antiferromagnetism. For
nearest neighbor interactions the TS-FCC lattice has an elegant
reinterpretation that is extremely useful. As shown in Fig. \ref{TorLat-b}
the honeycomb planes stack in such a way as to create
folded sheets of stacked triangular lattices.
The folded sheets run along two pairs of parallel edges in
the hexagon.  The remaining pair of edges bond neighbouring triangular
sheets.  This is made clear in Fig. \ref{TorLat-c} where we straighten out the triangular
sheets and draw the topologically equivalent semi-stacked triangular lattice or SSTL.
Unlike the case of the stacked triangular lattice (STL), in which each site has a nearest
neighbour in the sheets above and below it, the stacking bonds in the SSTL alternately
join sites in one sheet to the sheets above and
below.  The lattice co-ordination number is thus $7$ as it should be.

The TS-FCC lattice is one of an infinite subclass that share the same topology as the
SSTL: any
TS packing defined by stacking vectors that belong to one of the sets $\{{\bf V}_{\alpha i},
{\bf V}_{\beta i}\}$ is equivalent to triangular sheets stacked in this way. Thus
there are $3 \cdot 2^N$ such packings for $N$ layers---up to the overall factor of
$3$ for choice of sublattice diluted, this is the same as the number of the parent
Barlow packings.

\begin{figure}[ht!]
\begin{center}
\includegraphics[totalheight=9cm]{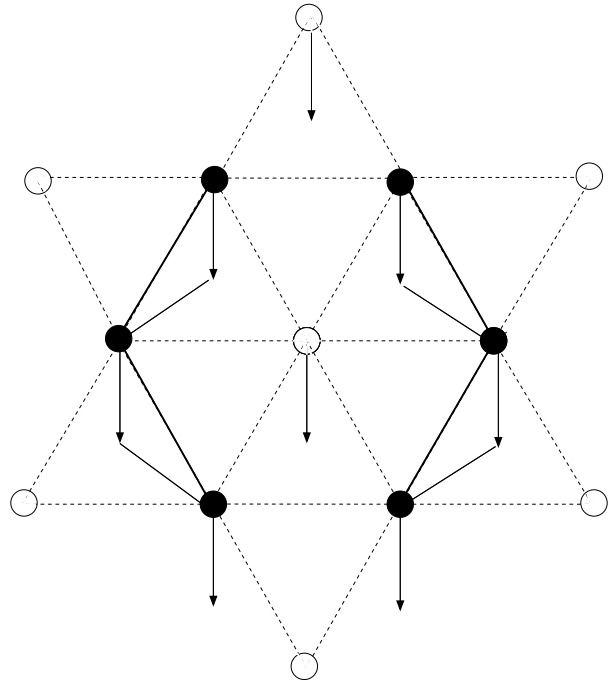}
\end{center}
\caption{\label{Tunnels} A schematic of the formation of triangular planes.  One layer
  of the parent triangular lattice is shown, with black (white) circles
  representing occupied (vacant) sites.  The arrows show the
  projection of the stacking vector $ V_{\alpha 2}$ in the honeycomb plane.  In a
  Barlow packing, the center of every upward facing triangle is
  an occupied site in the next layer, and the center of each triangle
  would be the apex of a tetrahedron.  Both solid and dotted lines are nearest neighbour
  bonds for the Barlow packing.  In the equivalent TS packing shown
  here, only the centers of the four triangles lying immediately below occupied sites
  are occupied in the next layer.  The solid lines show nearest neighbour bonds in the
  TS packing.  The darkened diagonal edges of the hexagon still
  form bases of triangles completed by the occupied sites in the next
  layer; the horizontal edges of the hexagon do not, and lie in the
  direction of stacking of the triangular sheets.  }
\end{figure}

To see how this comes about, let us begin with stacking one plane above a reference
plane with say ${\bf V}_{\alpha 2}$ and consider a given hexagon in the reference plane.
Let us label the three sets of parallel bonds on the hexagon by the indices on the
stacking vector projections orthogonal to them.  As we see in Fig~\ref{Tunnels}, two
of the three sets of parallel bonds on the hexagon are now also bonds on triangles
while one set---set 1---of parallel bonds is not. The same set is singled out when
we use stacking vector ${\bf V}_{\beta 2}$ instead.

It follows then that if we use a sequence of ${\bf V}_{\alpha 2}$ and
${\bf V}_{\beta 2}$ to stack,
we will get a sequence of honeycomb planes where the 2 and 3 bonds participate in
triangles and it is easy to convince oneself that this will lead to the claimed
topology. More precisely, the 2 and 3 bonds will lie in (folded) triangular planes
connected by 1 (stacking) bonds. Conversely, if we decide to switch from the 1 stacking
vectors to the 2 or 3 stacking vectors at some stage we will interfere with this
topology. Hence the result.

\begin{figure}[ht]
\begin{center}
\includegraphics[totalheight=5cm]{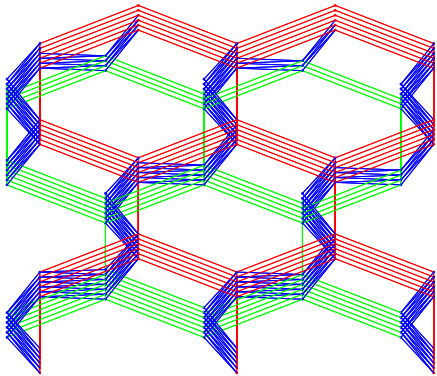}
\end{center}
\caption{\label{TSHCP} The TS-HCP lattice, showing stacking
structure.  Honeycomb planes are stacked according to an
alternating ABAB pattern.  The A planes are shown here in green
(dotted lines), and the B planes in red (solid lines).  In
contrast to the TS-FCC case, the tunnels formed by vacant sites
zig-zag between layers, giving the structure a 2 sublattice
chirality \cite{Torquato}.}
\end{figure}

We have already discussed the TS-FCC lattice obtained by repeated stacking with the
vector ${\bf V}_{\alpha 2}$. As another example we display, in Fig \ref{TSHCP}, the TS-HCP
structure constructed using the repeated sequence ${\bf V}_{\beta 2}, {\bf V}_{\alpha 2}$.
We emphasize that both of these have the topology of the SSTL in Fig.~\ref{TorLat-c}.

In the balance of this paper we will be concerned with $O(N)$ symmetric spins placed
on the sites of the TS-FCC lattice and other members of its class, interacting via
nearest-neighbor interactions alone. For these problems it will be sufficient to
consider such spins placed on the SSTL which is what we will do in the remaining.
This is a great simplification since it allows us to treat in one go an infinite
family of lattices with unit cells of arbitrarily large size. We will not treat the
problem of translating the results back to the original coordinates in the general
case except for the case of the TS-FCC lattice which we discuss in our concluding
remarks.

\section{Antiferromagnetism} \label{AF}

We now turn to nearest neighbor antiferromagnetism on the TS-FCC lattice
and its equivalents. As noted above, we will study the equivalent problems
on the SSTL.  Specifically, we wish
to elucidate the nature of ordering in the Hamiltonians,
\ba
H = \sum_{ij} J_{ij} S^a_i S^a_j \ ,
\ea
where $\sum_a S_i^a S_i^a =1$,  $a \in \{1,\cdots,N\}$, $i,j$ run over the
sites of the SSTL, and $J_{ij}=J$ when $i,j$ are nearest neighbors and zero
otherwise.
We begin by collecting some results on the eigenspectrum of the
nearest neighbor interaction (adjacency) matrix which will come in handy
in our subsequent analysis.

\subsection{Eigenspectrum of Interaction Matrix}  \label{BandStruct}

We wish to find the eigenvectors and eigenvalues of the adjacency matrix,
$J_{ij} \psi_j = \epsilon \psi_i$.
The SSTL differs from the STL in that translational symmetry is broken
along two of the triangular lattice vectors as well as along the stacking direction.
Consequently, it has a two site unit cell with sites of type $1$ connected only
to the triangular plane above, while sites
of type $2$ are connected only to the triangular plane below, as shown in Fig. \ref{TorLat-c}.
For convenience we switch to a co-ordinate system in which the triangular planes lie in the
$x-y$ plane, and stacking bonds in the $z$ direction.  With this choice the lattice vectors are
\ba
{\bf a_1} &=& a (1, 0, 0) \n
{\bf a_2} &=& a (1, -\sqrt{3},  0) \n
{\bf a_3} &=& a ({1 \over 2},- \frac{\sqrt{3}}{ 2},  1 )
\ea
and the 2-site unit cell now has sites at
\ba
{\bf u_0} &=& (0,0,0) \n
{\bf u_1} &=& a(\frac{1}{2}, -\frac{ \sqrt{3}}{2}, 0 ) \ .
\ea
Readers are warned not to mistake
this choice of axes for the SSTL for the choice of axes uses earlier to discuss
the TS-FCC lattice (Fig.  \ref{TorLat-a}); equally, the triangular planes
in the SSTL are not the triangular planes we began with in our discussion of the
parent Barlow packings.

The 2-site unit cell leads to eigenvectors that we parameterize in
the form
$$\psi_i \equiv \psi({\bf r}, \alpha) = e^{i {\bf k} \cdot {\bf r}}
u_\alpha ({\bf k}) \ ,$$
where ${\bf r}\equiv \{x,y,z\}$ is the actual location of the site of type
$\alpha=1,2$. The residual problem requires diagonalization of the
$2 \times 2$ reduction of the adjacency matrix to momentum space
\ba \label{Hsp}
\cos k_x \, I + [2 \cos{k_x \over 2} \cos{\sqrt{3} k_y \over 2}+\cos{k_z} ]
\, \sigma_x
 + \sin{k_z} \, \sigma_y  \nonumber
\ea

With these choices, the eigenvalues are
\ba
&\epsilon({\bf k}) / J &  =  \\
&\cos k_x & \pm  \left[  \sin^2 k_z
 + \left(2 \cos{k_x \over 2} \cos{ \sqrt{3} k_y \over 2} + \cos k_z
\right ) ^2 \right] ^{1/2} \nonumber
\ea
We will be especially interested in the minima of this dispersion relation as
they yield the soft modes that will dominate the ordering. Analysis of the
possible minima of $\epsilon({\bf k})/J$ reveals that $\epsilon_{min}/J = -2.5$,
and is attained for two inequivalent points in the Brillouin zone. We will,
however, find it convenient to choose two such points outside the first
Brillouin zone of the lattice as they facilitate comparison with the existing
analysis of the stacked triangular lattice. Accordingly, we will choose the pair:
\ba  \label{LowEPsis}
\psi_1 ({\bf r},\alpha) &=& e^{\frac{4 \pi i}{3}x} e^{i \pi z} \left( \begin{array}{c}
1 \\ 1 \\
\end{array} \right )  \n
\psi_2 ({\bf r},\alpha) &=& e^{-\frac{4 \pi i }{3} x} e^{-i \pi z}\left( \begin{array}{c}
1 \\ 1 \\
\end{array} \right )
\ea
Evidently, $\psi_2({\bf r}, \alpha) = \overline{\psi}_1({\bf r},\alpha)$.

\subsection{XY, Heisenberg and $N > 3$ cases} \label{XYAF}

For $N \ge 2$, which includes the XY and Heisenberg cases typically of
maximum interest, the ground states of the full lattice are simply the
well known coplanar, three sublattice ground states of the triangular
antiferromagnet, stacked antiferromagnetically between the different
layers. The reader will recall that the ground states of the triangular
antiferromagnet exhibit all spins confined to a plane in spin space with
three different orientations on the three sublattices making angles of
$120$ degrees with each other. There is a single global rotational
degree of freedom which carries over into the TS-FCC lattice. The set
of ground states is thus identical to those of the STL.

As these states thus involve breaking a continuous global symmetry
in three dimensions, we expect a single phase transition
between the paramagnetic phase at high temperatures and the $120$ degree
state at low temperatures. For the STL this transition has been discussed
extensively in the literature \cite{Kawamura} \cite{Kawa2} \cite{Loison}.
We will see that the results on the nature of the transition do not change
in our case although the details will of course be sensitive to the
altered microscopics.

\subsubsection{Landau-Ginzburg-Wilson functional} \label{GL1}

We will now follow the standard route of constructing the Landau-Ginzburg-Wilson
(LGW) functional
that controls the probability distribution of the soft modes from a
symmetry analysis. We will find that the LGW functional for the TS-FCC
lattice is essentially identical to that of the STL up to sixth order in the
fields and thus should be expected to lead to phase transitions in the same
universality class as the latter lattice.

We begin by writing (soft spin) configurations with energies near the two minima
(\ref{LowEPsis}) in the form:
\ba
\Phi^a ({\bf r}, \alpha) & = & \phi_1^a ({\bf r}) \psi_1 ({\bf r}, \alpha) +
\phi_2^a({\bf r}) \psi_2({\bf r}, \alpha) \n
& \equiv & \phi_1^a ({\bf r}) \psi_1 ({\bf r}, \alpha) +
\overline{\phi}_1^a({\bf r}) \overline{\psi}_1({\bf r}, \alpha)
\ea
where $a$ is the $O(N)$ vector index and on the second line we have built in the
real valuedness of the fields.

The reader can check that, of the various symmetry operations on the underlying
lattice, there are two that give independent non-trivial actions that need
to be considered in writing the LGW functional. These can be chosen to be a
translation by two steps in the $x$-direction,
\ba  \label{SymT1}
T_{x}^2 [ \phi_1^a ({\bf r}) ] &=& \phi_1^a ({\bf r} + 2 a \hat{\bf x} ) \n
& = & e^{\frac{2 \pi i}{3}} \phi_1^a ({\bf r})
\ea
and inversion,
\ba  \label{SymT2}
I[ \phi_1^a({\bf r})] &=& \phi_1^a(-{\bf r}) \n
 & = & \overline{\phi}_1^a({\bf r})
\ea
In addition, we must consider the $O(N)$ symmetry of the microscopic Hamiltonian.

Together these symmetries constrain the form of the LGW Hamiltonian to fourth
order in the fields to be:
\ba \label{GL}
\mathcal{H} =
&[& \! \! r+ c_\perp (q_x^2 + q_y^2) + c_z q_z^2 ] \, \phi_1^a \overline{\phi}_1^a   \n
&+& u_4 (\phi_1^a \overline{\phi}_1^a)^2+
v_4 (\phi_1^a \phi_1^a)(\overline{\phi}_1^b \overline{\phi}_1^b)
\ea
where we have summed over repeated indices. It is straightforward to confirm
that this Hamiltonian, when minimized, gives rise to the coplanar state we
deduce from the microscopic analysis.
As $\cal H$ has exactly the same form as for the STL and thus has been
studied extensively, we will now review the known results on its phase
transitions.

\subsubsection{Rernormalization group results on phase transitions}  \label{RG}

Renormalization group analyses of this Hamiltonian have been performed in the
literature both in the large $N$ and $d=4-\epsilon$ dimensional expansions
\cite{Kawamura88}.  An extensive review of these and other analytic and numerical
results can be found in Refs.~\onlinecite{Kawamura} and \onlinecite{Loison}.

This work has shown that there are four contending fixed
points, whose stability varies with $N$.  For $N>N_c$ there
is a single stable ``chiral'' fixed point, with $v_4 \ne 0$, which
controls a phase transition in a different universality class than that of
the ferromagnetic $O(N)$ model.

Depending on the initial parameters, the
flow may either lead to a second order transition at this
novel fixed point, or be unstable, signalling a first order
transition.  A simulation would be needed to settle this question for the
SSTL. For $N < N_c$ there are no stable fixed points,
and the transition is necessarily first order.

The most reliable estimate of $N_c$ comes from the Monte Carlo
Renormalization Group calculations of Ref.~\onlinecite{Itakura}.  These results
suggest that $4 < N_c <8$, and the cases of maximum physical interest lie in the
subcritical regime where the transition is first order.  This contradicts
the results of many earlier numerical studies, which seemed to indicate a
second order transition about the chiral fixed point.  The apparent
discrepancy stems from the presence of an attractive basin in the flow
about complex
fixed points lying close to the real plane, which causes the transition to
appear second order for small system sizes \cite{Loison}.

\subsection{Ising case}  \label{Ising}

Thus far our analysis of the TS-FCC lattice has closely paralleled
the analysis of the STL. But now for the Ising case, a new and
interesting feature enters which distinguishes the two lattices.
As is well known, a single triangular Ising layer exhibits a
macroscopic number of ground states \cite{wannier, Houtappel}. In
the STL the ground states of the stacked lattice are as many since
they consist of single layer ground states repeated
antiferromagnetically (although the number can be boosted somewhat by
picking antiperiodic boundary conditions in the stacking direction).
For a three dimensional system, this is a
submacroscopic number of states and thus the entropy per site
vanishes as $T \rightarrow 0$. For the SSTL we find instead that the
number of ground states is again macroscopic and now there is a
non-zero entropy per site as  $T \rightarrow 0$.

Despite this difference, the nature of the ordering at low temperatures
in both systems---driven by the order by disorder mechanism---turns out
to be the same. This is indicated by the coincidence of their LGW functionals
(up to coefficients) and we are also able to give numerical and analytic
evidence to the same end.

\subsubsection{Zero temperature entropy}

\begin{figure}[htp]
\begin{center} \label{Maxflip}
\includegraphics[totalheight=5cm]{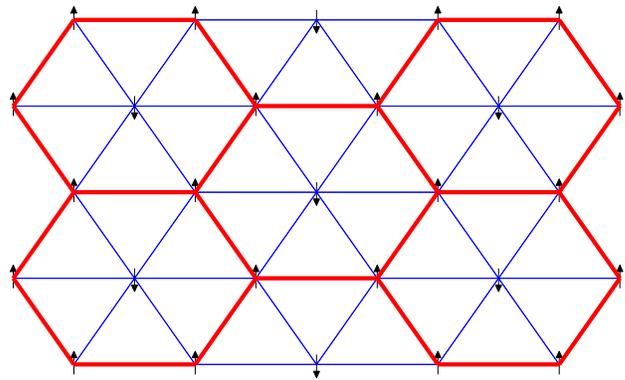}
\caption{Maximally flippable configuration.  Ising spins are shown
on each site. Frustrated bonds are (bold) red, unfrustrated bonds
(light) blue.}
\end{center}
\end{figure}

Let us first consider a lower bound on the zero temperature entropy.
We begin with the ``maximally flippable configuration'' in a single triangular
plane shown in Fig \ref{Maxflip}. In this configuration, spins on two out of three
sublattices  are flippable, in that they can be individually flipped
without leaving the ground state manifold.
This configuration has as many flippable spins as can be packed into a
ground state. Observe that sites on one of the
two sublattices are independently flippable and thus generate $2^{N/3}$
ground states that bound the entropy
of an isolated plane from below by $(\log{2})/3$ per site.

Now consider stacking this configuration antiferromagnetically.
In a given plane, half of the sites are married to sites in
the layer below, and the other half above. It follows then that we
may flip half the sites on one sublattice along with their partners
above and the other half with their partners below. This leads to a
lower bound on the ground state entropy
\be \label{ST0}
S/N > (\log 2)/6
\ee
where $N$ is the total number of sites in the system. The scaling with
$N$ establishes the macroscopic character of the ground state entropy.
In contrast, for the STL there are only $N^{2/3}$ ground states. A simple
upper bound
\be
S_u/N < 0.3383 \ldots
\ee
is obtained by considering the entropy of decoupled triangular layers \cite{wannier}.

We remark that a binary alloy that forms in the TS-FCC family of structures
would thus be expected to exhibit a macroscopic zero temperature entropy,
contributing to its stabilization.

\subsubsection{Order by disorder} \label{obd}

The next question to consider is whether the ground state manifold
breaks any symmetries, i.e. whether the unweighted average over
all the ground states yields long range order in the correlation functions.

What kind of order might one expect? As this order has to be
selected entropically, i.e. by the preponderance of a family of
configurations in the ground state average, we expect it to
correspond to the configuration that has the greatest number of
nearby configurations reached by local moves. The stacked
maximally flippable (MF) configuration considered in our entropy
lower bound meets this criterion---it is also the three
dimensional configuration with maximal flippability.  To see this,
observe first that the constraint of inter-planar spin partnering
is absolute in the ground state manifold: no spin may be flipped
independently of its partner. Spin configurations which are {\it
stacked} (the same in every layer) automatically partner flippable
spins to flippable spins, and thus the stacking bonds impose no
additional constraints on flippability.  As the MF state maximizes
the number of flippable spins in each plane, stacking this state
gives the maximum possible number of flippable spins for the SSTL.

We should note however, that the spin distribution in the maximally
flippable configuration is not directly observable; instead, it
must be dressed by the fluctuations that select it. Two options
emerge naturally.  The first involves a three sublattice
structure with magnetizations $(c,-c,0)$ wherein one of the two
sublattices of flippable spins does all the flipping and thus
exhibits a vanishing magnetization while the other two sublattices
exhibit equal and opposite magnetizations. The other exhibits a
three sublattice structure but now with two equivalent
sublattices. The magnetizations $(d, -d'/2 -d'/2)$ reflect more
completely the symmetries of the maximally flippable
configuration. The selection between these two is a matter of
detail. The reader should note that both options give rise to
six symmetry equivalent states.

Unfortunately, direct demonstration that one of these options
is realized is not straightforward and we will not definitively
answer this question in this paper although we believe that
symmetry breaking in the $(c,-c,0)$ is realized at $T=0$. Instead
we will, in the next section, approach the existence and structure
of the ordered phase from the paramagnetic phase at high temperatures
by constructing the appropriate LGW functional.

But before we do that let us briefly comment on the difference
between what we have discussed here the corresponding analysis of
the STL Ising antiferromagnet. On the STL, the ground state
manifold exhibits long range order in the stacking direction but
only algebraic order in the planes---in the latter directions it
exhibits the known correlations of a single triangular layer
\cite{stephenson}. This algebraic order is again present at the
wavevectors of the maximally flippable state (Fig \ref{Maxflip}).
In the STL, switching on a small $T >0$ converts this to true long
range order. The mechanism is ``order by disorder'' which can be
visualized as the entropic dominance of three dimensional
configurations in which flippable spins in the MF configurations
in the planes stack with a set of mobile solitonic defects
\cite{coppersmith,mcs2000,ms2001}. In this setting it is by now
clear that a single low temperature phase in the $(c,-c,0)$
pattern is separated from the paramagnet \cite{kim1990,mcs2001}.
The major qualitative difference between the STL and the SSTL is
then that in the latter fluctuations in the stacking direction are
present already at $T=0$ and so we expect that (eventually) the
low temperature ordering can be understood by an analysis of the
ground states alone.

\subsubsection{LGW analysis}   \label{GL2}

We now add another ingredient to our analysis of the Ising problem
by applying the LGW and Renormalization group analysis to this
case. This yields \ba \label{GLIsing} \mathcal{H_I} =
&[& \! \! r+ c_\perp (q_x^2 + q_y^2) + c_z q_z^2 ] \, \phi_1^a \overline{\phi}_1   \\
&+& u_4 (\phi_1 \overline{\phi}_1)^2+
 u_6 (\phi_1 \overline{\phi}_1)^3  + v_6 (\phi_1^6 + \overline{\phi}_1^6)
\nonumber \ea where we have now kept terms to sixth order in the
fields. This is necessary for the second of these terms is the
first one that breaks a $U(1)$/XY symmetry that is present up to
fourth order down to a $Z_6$ (clock) symmetry. Consequently, there
is a discrete set of six symmetry equivalent states at low
temperatures and we reproduce a key feature of the Ising problem.
The two possible signs of $v_6$ correspond to the two
magnetization patterns discussed in above. This term is
dangerously irrelevant: it is irrelevant at the critical fixed
point that controls the transition into the broken symmetry phase,
but to get the correct low-temperature physics it cannot be set to
zero. Since it {\it is} irrelevant at the critical point, the
transition is in the universality class of the three dimensional
XY model. It is worth noting that a finite stack will exhibit a
Kosterlitz-Thouless transition \cite{RM-KT}. All these results
parallel those for the STL \cite{Blankschtein}.

\subsubsection{Monte Carlo results}

The remaining challenge is to distinguish between the two ordering
alternatives or equivalently, to fix the sign of $v_6$. As this is
sensitive to microscopics, we have chosen to simulate the
system to investigate this question.

In the simulations we used a simple spin-flip Monte Carlo
algorithm. The algorithm allows 2 types of moves: a single spin
flip, or a double spin flip which reverses a pair of partnered
spins in adjacent layers. While this is sufficient for our
purposes near the transition, at low temperatures it fails to be
ergodic.  The nature of the problem is clearest at zero
temperature, where only the double spin flip is allowed. Hence a
spin $s_1$ in a given triangular plane may be flipped only if its
partner $s_2$ in the adjacent plane is also flippable.  As only
$2$ of the $6$ in-plane neighbours of $s_1$ are partnered with
in-plane neighbours of $s_2$, at $T=0$ configurations exist in
which this pair can be flipped only after flipping spins in all
other layers of the system.  Hence at low temperatures a more
complicated, cluster-type method must be used.  We expect to
discuss such a method, and thus the ordering at low temperatures
as well as a good estimate of the zero temperature entropy in a
future publication \cite{wip}.

We turn now to the data for systems of size $6 \times 6 \times 6$ and
$ 12 \times 12 \times 12$ for the temperature ranges where our algorithm
is ergodic as evidenced by the decay in single spin autocorrelations to
zero. The system dimensions are chosen to be $N$ triangular planes of
$N^2$ sites each with periodic boundary conditions in all directions.

\begin{figure}[htp]
\begin{center} \label{MagCors}
\includegraphics[totalheight=5cm]{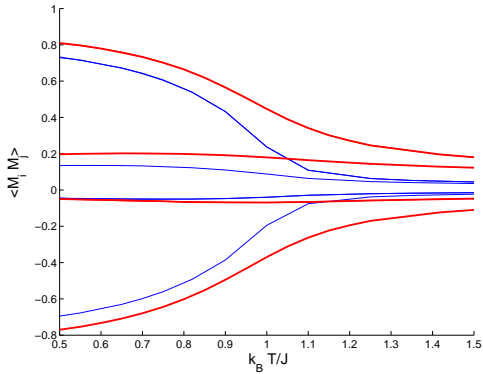}
\caption{Correlations of sublattice magnetizations as functions of
temperature, indicating the phase transition from 3 sublattice order to a
paramagnetic state.  Two lattice sizes are shown: a $6 \times 6 \times 6$
lattice in (light) blue, and a $12 \times 12 \times 12$ lattice in (bold) red.}
\end{center}
\end{figure}

For these systems we proceed as follows: In each configuration
we compute and order the three sublattice magnetizations as $M_1 >
M_2 > M_3$. We then compute the matrix of correlations $\langle
M_i M_j \rangle$ averaged over the run. The results are shown in
Fig~\ref{MagCors}. As the reader will note these correlators should,
in the $(c,-c,0)$ state, exhibit the values $c^2$, $-c^2$ and $0$ in the
infinite volume limit. Our computations are consistent with that and
clearly inconsistent with the competing  $(d, -d'/2 -d'/2)$ state.
This includes details such as the multiplicity of the values observed,
and their evolution between the two system sizes.

\section{Concluding Remarks} \label{Conclusion}

To summarize, we have studied nearest neighbor $O(N)$ antiferromagnets on
an infinite subset of the new family of packings introduced by Torquato
and Stillinger and established the nature of the ordering at low temperatures
and the nature of the phase transitions.

We have done our analysis in the equivalent representation of the
SSTL. This has the great advantage that we have dealt with the
entire family of lattices at once---most of which have sizeable
unit cells stemming from long periods in the stacking direction.
However, for a specific realization, it will be necessary to
translate the ordering back into the actual geometry of the
lattice. For example, for the TS-FCC lattice the ordering
wavevectors $\pm{4 \pi \over 3} \hat{x}$ common to all $O(N)$
cases,  translate into the vectors
$$\pm {2 \pi \over a} \left( 0, {\sqrt{3} \over 9}, {\sqrt{6} \over 9} \right )$$
in the choices made in Equations (\ref{primlatvecs}) and (\ref{basisvecs}).

One statistical mechanical remark may be interesting to readers. By this somewhat
circular route we have discovered that the SSTL preserves the ordering of the
STL for the Ising problem, while exhibiting a greatly increased ground state
entropy. This analysis indicates that further dilution of the stacking bonds
will further boost the zero temperature entropy while still preserving the nature
of the ordered phase at asymptotically low temperatures.

Finally, we have taken a preliminary look at TS packings which are not in the
TS-FCC class. The appear, generically, to be more frustrated than the ones
studied in this paper and thus are an interesting topic for future work.

\suhe{Acknowledgements}
We are very grateful to Sal Torquato for introducing us to the
Torquato-Stillinger packings. We would also like to thank Werner
Krauth and Roderich Moessner for discussions on cluster and pocket
algorithms. This work was supported in part by NSF Grant No. DMR
0213706 (SLS) and a NSERC fellowship (FJB).

\bibliography{TSBib.bib}

\end{document}